\documentclass[11pt]{article}
\usepackage{geometry}                
\usepackage[small]{caption}
\usepackage{subfig}
\usepackage{graphicx}
\usepackage{amsmath}
\usepackage{amsfonts}
\usepackage{amssymb}
\usepackage{epstopdf}
\usepackage{mathrsfs}
\usepackage[T1]{fontenc}
\usepackage{feynmf}
\title{The N-Quantum Approximation and Bound States in Motion}
\author{Steve Cowen\thanks{
e-mail address, scowen@umd.edu} \\
\emph{Center for Fundamental Physics}\\
\emph{University of Maryland} \\
\emph{College Park, MD 20742, US} \\
University of Maryland Preprint PP-11-010}

\begin{document}

\maketitle

\begin{abstract} 
We use an alternative method to the Bethe-Salpeter equation, the N-Quantum approximation (NQA), for studying bound states in motion.  We use this method to find a relativistic equation for weakly bound states of two constituents with different masses.  We present rules for interpreting simple diagrams associated with the NQA.  We can use these rules to construct the bound state equations directly, avoiding some of the complications of the process.   The final result is a bound state equation that shows Lorentz contraction in the direction of motion explicitly.  This result matches that of \cite{Jarvinen} found using the Bethe-Salpeter equation.  We briefly discuss some other applications of the NQA in studying the effects of motion on bound states.
\end{abstract}

\section{Introduction}

The purpose of this paper is to show how the N-Quantum procedure can be used to study bound states in motion.  While our results are similar to those found using the Bethe-Salpeter equation, the procedures are quite different.  The effects of motion on a bound state have been analyzed within the framework of the Bethe-Salpeter approach 
in \cite{Kim}, \cite{Jarvinen1}, and \cite{Jarvinen}.  While the success of the equation in describing bound state phenomena is undeniable, 
it is not without its shortcomings.  As pointed out in \cite{green}, \cite{be1}, and \cite{be2}, the B-S method suffers from spurious solutions 
with negative norm amplitudes.  It is also difficult to interpret the relative time coordinate of the B-S equation.  The NQA 
avoids these problems and may also be easier to employ in certain situations. 

The main idea of the N-Quantum approach is to expand the interacting fields that appear in the Lagrangian or Hamiltonian in terms of \emph{in} fields.  
These \emph{in} fields are related to eigenstates of the Hamiltonian with quantum numbers of freely moving asymptotic incoming particles.  We assume
that these fields form a complete set. The Haag expansion of the interacting fields in terms of \emph{in} fields is generally an infinite series.  
For the purpose of approximation, we terminate this series, keeping only terms with a small number of \emph{in} fields.  Each 
term in the series contains an undetermined function of the relevant coordinates known as a Haag amplitude.  The goal of the NQA is to derive 
an equation, or a set of equations, that can be used to solve for these amplitudes.  We accomplish this by taking the equations of motion for 
the interacting fields, expanding each of the fields in normal-ordered products of $in$ fields, and renormal ordering.  We remove residual \emph{in} fields by contracting 
with external \emph{in} fields.  After all \emph{in} fields are contracted, the results are bound state equations for the amplitudes.  If only 
low order terms are used in the Haag expansions, these equations will be linear in the amplitudes.  

Although they sometimes arrive at similar conclusions, the N-Quantum procedure is quite different than the Bethe-Salpeter method.
  The derivation of the Bethe-Salpeter equation begins with the Dyson equation for the two particle Green function, while the N-Quantum's
 roots are in Haag's operator expansion.   Unlike the Bethe-Salpeter wave function, the position space Haag amplitudes depend only on three-vectors.  With the use of these amplitudes, the N-Quantum procedure avoids using the relative time coordinate while still maintaining a covariant formalism.  The Haag amplitude is similar to a Bethe-Salpeter amplitude with one of the constituent's mass shell singularity removed and that constituent's momentum restricted to the mass shell.

The first section of this paper reviews the NQA for a relativistic bound state composed of two fermions in motion.  A similar model for a bound state at rest composed of two scalars mediated by a third scalar was analyzed in \cite{green2}.  Other applications of the NQA, such as the study of symmetry breaking, scaling limits, and a deuteron model, can be found in \cite{green3}, \cite{green4}, and \cite{green5}.  Although working in position space is in some ways more intuitive, we work in momentum space in this paper to make the calculations simpler.  We describe the standard N-Quantum procedure and show how to draw and interpret diagrams associated with the bound state equation.  These diagrams are similar to Feynman diagrams, but must take into account difference between off-shell and on-shell lines.  We develop a set of rules for interpreting the diagrams to facilitate future calculations of more complicated diagrams.  

Section (3) begins with a change of variables to relative and total momentum.  After expanding certain factors in these new variables and using some approximations proposed in \cite{Jarvinen}, we rewrite the equation of motion to explicitly display the Lorentz contraction of the amplitude.  Our results for the case with different masses for the two particles reduce to the equal mass case of \cite {Jarvinen}.  We plan on discussing higher Fock state contributions in a future paper.

\section{The N-Quantum Procedure}
We refer to fermion 1 as an electron and fermion 2 as a proton.
We begin with the equations of motion.  The electron momentum space equation of motion (neglecting weak interactions) is
\begin{align}
(\gamma^\mu p_\mu-m_1)\psi_1(p)&=\frac{e\gamma^\mu}{(2\pi)^{3/2}}\int d q dk \delta(p-k-q) A_\mu(q)\psi_1(k)
\end{align}
and the photon equation in Feynman gauge is
\begin{align}
-p^2 A^\mu(p)&=\frac{e}{(2\pi)^{3/2}}\int dk_1 dk_2 \delta(p+k_1-k_2) \bar{\psi}_s(-k_1)\gamma^\mu \psi_s(k_2) 
\end{align}
where we have defined $\bar{\psi}(p)=\psi(p)^\dagger \gamma^0$ and the sum over the subscript $s=1,2$ is implied.  Using Eq. (2) we can rewrite Eq. (1) as
\begin{align}
(\gamma^\mu p_\mu-m_1)\psi_1(p)&=-\frac{e^2\gamma^\mu}{(2\pi)^3}\int d k dl_1 dl_2 \frac{\delta(p-k-l_2+l_1)\bar{\psi}_s(-l_1)\gamma_\mu \psi_s(l_2)\psi_1(k)}{(l_2-l_1)^2}
\end{align}
Since we will be substituting expansions of the interacting fields, it is important to properly symmetrize these fields.  This is also way of imposing charge conjugation symmetry.  After symmetrizing, the equations become
\begin{align}
-p^2 A^\mu(p)&=\frac{e}{2(2\pi)^{3/2}}\int dk_1 dk_2 \delta(p+k_1-k_2) [\bar{\psi}_s(-k_1),\gamma^\mu \psi_s(k_2)]_- 
\end{align}
\begin{align}
(\gamma^\mu p_\mu-m_1)\psi_1(p)&=-\frac{e^2\gamma^\mu}{ 4(2\pi)^3}\int d k dl_1 dl_2 \frac{\delta(p-k-l_2+l_1)[[\bar{\psi}_s(-l_1),\gamma_\mu \psi_s(l_2)]_{-},\psi_1(k)]_{+}}{(l_2-l_1)^2}
\end{align}

We now expand the interacting fields in terms of \emph{in} fields.  Although only a few of the terms in expansions are relevant, we show some other terms that may be useful in other calculations as well.  The Haag expansions are
\begin{align}
\psi_1(p)&=\psi_1^{in}(p)+\int d^4r d^4b \delta(p+r-b)  \mathscr{F}_1(r,b) :\bar{\psi}_2^{in}(-r) B^{in}(b): \notag \\ 
& \;\;\;\;  +\int d^4r d^4q d^4b \delta(p+r-q-b)  \mathscr{H}_1^\mu(r,b):A_\mu^{in}(q)\bar{\psi}_2^{in}(-r) B^{in}(b):  \\ 
A^\mu(p)&=\int d^4r_1 d^4r_2 \delta(p+r_1-r_2) :\bar{\psi}_2^{in}(-r_1)\mathscr{G}^\mu(r_1,r_2) \psi_2^{in}(r_2): \notag \\
& \;\;\;\; +\int d^4b d^4r_1 d^4r_2 \delta(p+r_1+r_2-b) \mathscr{J}^\mu(r_1,r_2,b) :\bar{\psi}_2^{in}(-r_2) \bar{\psi}_1^{in}(-r_1)B^{in}(b): 
\end{align}
where $B^{in}(b)$ is the bound state field, the colons represent normal ordering, and spinor indices are suppressed.  The expansion for the proton field, $\psi_2$, is found by interchanging $1\leftrightarrow 2$ in the electron equation.

Contractions between two momentum space $in$ fields are
\begin{align*}
<0| \psi^{in}_{1\alpha}(p_1) \bar{\psi}_{1\beta}^{in}(p_2)|0>&=(\displaystyle{\not} p_1+m_1)_{\alpha \beta}\delta^{(+)}_{m_1}(p_1)\delta^4(p_1+p_2) \\
<0| \bar{\psi}_{1\beta}^{in}(p_2) \psi^{in}_{1\alpha}(p_1) |0>&=-(\displaystyle{\not} p_1+m_1)_{\alpha \beta}\delta^{(-)}_{m_1}(p_1)\delta^4(p_1+p_2)\\
<0| A^\mu(p_1) A^\nu(p_2)|0>&=-g^{\mu \nu} \delta^{(+)}_0(p_1)\delta^4(p_1+p_2)
\end{align*}
where $\delta^{\pm}_{m}(p)=\theta(\pm p^0)\delta(p^2-m^2)$.  We insert the Haag expansions for the electron and proton fields into the right hand side of Eq. (5) and contract to find the two main terms that contain the factor $\bar{\psi}_2^{in} B^{in}$. (There are two additional terms that are used in renormalization.)  We "peel" these fields off by contracting with an external $\psi_2^{in}$ and $B^{in}$ field.  This is equivalent to replacing the fields with a factor of $(\displaystyle{\not} p+m_2)$, where p is the momentum of the proton, and dropping the integrals over their momenta.  The result is
\begin{align}
RHS=T_1+T_2 
\end{align}
where
\begin{align*}
T_1&=-\frac{e^2(\gamma^\mu)}{(2\pi)^3}\delta_{m_2}(p_1)\delta(p-b+p_1) \int d^4l_2  \mathscr{F}_1(l_2,b)[(\displaystyle{\not} l_2+m_2)\gamma_\mu(\displaystyle{\not} p_1+m_2)]^T  \notag\\
&\;\;\;\;  \times \frac{\delta_{m_2}(l_2)}{(l_2-p_1)^2}
\end{align*}
and
\begin{align*}
T_2&=\frac{ e^2(\gamma^\mu)}{(2\pi)^3}\delta(p-b+p_1)\delta_{m_2}(p_1) \int  d^4k  (\displaystyle{\not} k+m_1) [(\displaystyle{\not} p_1+m_2)(\gamma_\mu) \mathscr{F}_{2}(k,b)]^T   \notag\\
&\;\;\;\;  \times \frac{\delta_{m_1}(k)}{(b-k-p_1)^2}.
\end{align*}
Defining $f_1(p,b)\equiv\mathscr{F}_1(p,b)(\displaystyle{\not} p+m)^T C$ and using $C(\gamma^\mu)^T C = -\gamma^\mu$, $C=\gamma^0 \gamma^2$, we can write the bound state equation in the tidier form
\begin{align}
(\displaystyle{\not} b-\displaystyle{\not} p-m_1) f_1(p,b)&= \frac{e^2}{2(2\pi)^3}\int  d^4p' \gamma^\mu \bigg\{ \frac{\delta_{m_2}(p') f_1(p',b)}{k^2} \notag \\
 &\;\;\;\; -\frac{\delta_{m_1}(p') C^T f_{2}(p',b)^TC}{{k'}^2}\bigg\} \gamma_\mu(-\displaystyle{\not} p+m_2),
\end{align}
where $k=p-p'$ and $k'=b-p-p'$.

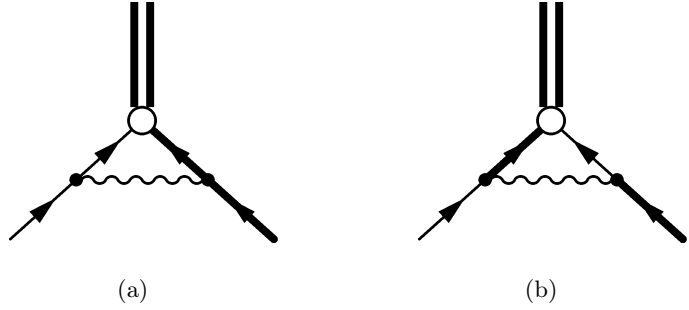
\begin{figure}[t]
\begin{center}
\subfloat[]{\label{fig:a}
\begin{fmffile}{fd}
\begin{fmfgraph*}(100,100) \fmfpen{thin}
\fmfbottomn{i}{2} \fmftop{o1} 
\fmf{fermion,label=$b-p$}{i1,v1}
\fmf{fermion}{v1,v3}
\fmf{photon,tension=0}{v1,v2}
\fmf{fermion,width=25,label=$p$}{i2,v2}
\fmf{fermion,width=25}{v2,v3}
\fmf{double,width=25,label=$b$}{v3,o1}
\fmfdot{v1,v2}
\fmfv{decor.shape=circle,decor.filled=empty,
decor.size=5thick}{v3}
\end{fmfgraph*}
\end{fmffile}}
\;\;\;\;\;\;\;\;\;\;\;\;
\subfloat[]{\label{fig:b}
\begin{fmffile}{diagram2}
\begin{fmfgraph*}(100,100) \fmfpen{thin}
\fmfbottomn{i}{2} \fmftop{o1} 
\fmf{fermion,label=$b-p$}{i1,v1}
\fmf{photon,tension=0}{v1,v2}
\fmf{fermion,width=25}{v1,v3}
\fmf{fermion,width=25,label=$p$}{i2,v2}
\fmf{fermion}{v2,v3}
\fmf{double,width=25,label=$b$}{v3,o1}
\fmfdot{v1,v2}
\fmfv{decor.shape=circle,decor.filled=empty,
decor.size=5thick}{v3}
\end{fmfgraph*}
\end{fmffile}}

\end{center}
\caption{Graphs for the right hand side of the electron equation of motion.  Heavy lines are on shell and light lines are off shell.  The heavy double line represents the bound state (hydrogen atom).  The empty circle represents the amplitude $F_1$ in (a) and $F_2$ in (b).  The left fermion line is the electron and the right line is the proton.  Similar graphs exist for the proton equation.}
\end{figure}

We have arrived at the desired bound state equation through a somewhat arduous procedure.  It would be useful for future calculations to develop a process for drawing and interpreting graphs, rather than contracting fields and simplifying more complicated expressions.  The diagrams that are relevant to the preceding calculation must have one external off-shell electron line, an external on shell proton line, and an external on-shell bound state line.  The two possible lowest order diagrams are shown in Figure 1.  The rules for analyzing the diagrams associated with the N-Quantum procedure are similar to Feynman rules, but they must also accommodate on-shell lines.  The rules are:
\begin{enumerate}
\item Draw all possible relevant low order diagrams.  In this case, these are diagrams with an "incoming" off-shell electron and on-shell proton and an "outgoing" on-shell bound state.  Amplitude vertices should be distinct from normal vertices.  Care must be taken to ensure the correct ordering of the following factors.
\item Write a factor of $\frac{1}{\displaystyle{\not} p -m}$ for any off-shell line not connected an amplitude, where $p$ is the momentum of the line and $m$ is its mass (This factor is on the left hand side in Eq. (9)).
\item Write $e\gamma^\mu$ for every fermion-photon vertex.
\item Write nothing for any off shell line connected to an amplitude.
\item Write a factor of $\delta_m(p')f_i(p',b)$ for the bound state vertex, where $p'$ is the 4-momentum of the on shell fermion line connected to the bound state vertex and $m$ is its mass.  Take a transpose, then left multiply by $iC^T$ and right multiply by $iC$ if the internal off shell fermion line is different from the external off shell line.\footnotemark
\item Write $-g_{\mu \nu}/k^2$ for every internal off shell photon line, where k is the momentum of the photon line.
\item Integrate over the internal on shell momentum with a factor of $(2\pi)^{-3}$.
\item Add a factor of $(-\displaystyle{\not} p+m)$ for any on shell external fermion line.
\item Add a symmetry factor, in this case, $\frac{1}{2}.$
\end{enumerate}
These rules must be slightly revised for more complicated diagrams, such as those including off shell internal fermion lines not connected to any amplitude.  

 \footnotetext{The full mass shell function $\delta_m(p')$ is a result of symmetrization and specific to this example.  For higher order terms, expressions will exist with higher order factors of $\delta^{(\pm)}_m(p)$.  Symmetrization creates specific functions of the mass shell delta-functions such as $\Theta_1(p_1,p_2)$ found in Eq. (18).}

A noticeable feature of Eq. (9) is the inclusion of both positive and negative mass shells on the right hand side.  The opposite mass shell was dropped in \cite{Ray}, a paper that uses the N-Quantum to study the hydrogen atom at rest.  It is easy to see why the opposite mass shell is negligible.  Looking at the $k^2$ in the denominator, we find when the momentum $p$ is on the opposite mass shell as $p'$
\begin{eqnarray*}
k^2&=&(E_p+E_{p'})^2-(\mathbf{p-p'})^2 \\
&\approx&[\kappa_2 \epsilon \left(1+\frac{(\frac{\mathbf{q}}{\kappa_2})^2-\frac{2|\mathbf{b}|q_\parallel}{\kappa_2}}{2 \epsilon^2}\right)+\kappa_2\epsilon \left(1+\frac{(\frac{\mathbf{q'}}{\kappa_2})^2-\frac{2|\mathbf{b}|{q_\parallel}'}{\kappa_2}}{2 \epsilon^2}\right)]^2 -(\mathbf{p-p'})^2 \\
&=& (2\kappa_2 \epsilon)^2 +O(\alpha)
\end{eqnarray*}
where $\epsilon=\sqrt{\mathbf{b}^2+(m_1+m_2)^2}$, $\kappa_2=\frac{m_2}{m_1+m_2}$ and $\mathbf{q}$ is the relative momentum defined in the following section.  When the two momenta are on the same mass shell, we find
\begin{eqnarray*}
k^2 \approx-(\mathbf{k}_\perp^2+\gamma^{-2}k_\parallel^2) \sim O(\alpha^2),
\end{eqnarray*}
a result found in the following section.  The opposite mass shell term is suppressed by a factor of order $O(\frac{\alpha^2}{\epsilon^2})$ relative to the other mass shell.  While the negative mass shell does seem to be small compared to the positive shell, it may be important when analyzing higher order contributions.

We can show that the second term on the RHS of Eq. (9) is equal to the first term to lowest order by finding a relation between the amplitudes $F_1$ and $F_2$.  We find this relation by starting with the equal time anticommutator 
\begin{align*}
[\psi_1(\mathbf{x},t),\psi_2(\mathbf{y},t)]_+=0,
\end{align*}
Fourier transforming, Haag expanding, and contracting \emph{in} fields.  The relation between the two momentum space amplitudes is
\begin{align}
\delta_{m_2}(l)f_1(l,b)=-\delta_{m_1}(p)C^Tf_1(p,b)^T C.
\end{align}
where $l=b-p$.  Using this, the second term becomes
\begin{align}
T_2&=- \frac{e^2}{2(2\pi)^3}\int  d^4p' \gamma^\mu \frac{\delta_{m_1}(p') C^T f_{2}(p',b)^TC}{{k'}^2} \gamma_\mu(-\displaystyle{\not} p+m_2) \notag \\
&= \frac{e^2}{2(2\pi)^3}\int  d^4p' \gamma^\mu \frac{\delta_{m_2}(b-p')f_1(b-p',b)}{{k'}^2} \gamma_\mu(-\displaystyle{\not} p+m_2)\notag \\
 &= \frac{e^2}{2(2\pi)^3}\int  d^4p' \gamma^\mu \frac{\delta_{m_2}(p')f_1(p',b)}{{k}^2} \gamma_\mu(-\displaystyle{\not} p+m_2)\notag
\end{align}
This term exactly matches the first term.  It should be noted that this approximation used only lower order terms in the Haag expansion.  We could have kept the bound state equations as a set of coupled equations and solved them numerically for a more exact result.

\section{Hydrogen in motion}
In this section, our goal is to arrive at an equation that shows the Lorentz contraction of a moving bound state.   This equation will be similar to that found in \cite{Jarvinen}, but with different masses for the constituents.  Although the results resemble each other, the derivations are quite different.  We must first define the relative momentum by 
\begin{align}
q&\equiv p-\kappa_2 b
\end{align}
where $\kappa_i=\frac{m_i}{M}$ and $M=m_1+m_2$.  In terms of the relative momentum, the proton has momentum $q+\kappa_2 b$ and the electron has momentum $b-p=-q+\kappa_1 b$.  We use the notation $E^{(i)}_p=\sqrt{\mathbf{p}^2+m_i^2}$ to distinguish between the energies of the two particles, and the subscripts $\parallel$ and $\perp$ to indicate components parallel and perpendicular to the bound state motion.  

We begin with Eq. (9) with the approximation of Eq. (10).  We can left multiply by the inverse of the kinetic energy operator on both sides to get
\begin{align}
 f_1(p,b)=  \frac{e^2}{(2\pi)^3}\frac{(\displaystyle{\not} b-\displaystyle{\not} p+m_1)}{(b-p)^2-m_1^2} \int  d^4p' \gamma^\mu f_1(p',b)\delta_{m_2}(p') \gamma_\mu\left(\frac{1}{k^2}\right)(-\displaystyle{\not} p+m_2).
\end{align}
Manipulating the denominator outside of the integral, we find
\begin{align}
(b-p)^2-m_1^2 &= (E-E^{(2)}_{p})^2-(\mathbf{b-p})^2 -m_1^2 \notag \\
&= \epsilon^2+ 2 \Delta E (\epsilon-E^{(2)}_{\mathbf{q}+\kappa_2 \mathbf{b}})-2 \epsilon E^{(2)}_{\mathbf{q}+\kappa_2 \mathbf{b}} -\mathbf{b}^2+2 \mathbf{b}\cdot (\mathbf{q}+\kappa_2 \mathbf{b})\notag\\
& \;\;\;\;+M^2 (\kappa_2-\kappa_1)+O(\alpha^3) \notag \\
&=\epsilon^2+2 \epsilon \Delta E(1-\kappa_2)-2 \epsilon [\kappa_2 \epsilon \left(1+\frac{\left(\mathbf{\frac{q}{\kappa_2}}\right)^2+\frac{2b q_\parallel}{\kappa_2}}{2 \epsilon^2}-\frac{1}{8}\left(\frac{\frac{2b q_\parallel}{\kappa_2}}{\epsilon^2}\right)^2\right)]\notag \\
& \;\;\;\;+2 \mathbf{b\cdot q}- (\mathbf{b}^2 +M^2) (\kappa_1-\kappa_2)+O(\alpha^3) \notag\\
&=-\frac{1}{\kappa_2}(\mathbf{q}_\perp^2+\gamma^{-2} q_\parallel^2-2 \kappa_1 \kappa_2\Delta E \epsilon)
\end{align}
where $\Delta E=E-\epsilon$ , $\epsilon=\sqrt{\mathbf{b}^2+M^2}$ , $\mathbf{\beta}=\frac{\mathbf{b}}{\epsilon}$ , and $\gamma=(1-\beta^2)^{-1/2}$.
The denominator of the integrand is
\begin{align*}
k^2={k^0}^2-\mathbf{k}^2={k^0}^2-\mathbf{k}_\perp^2-k_\parallel^2.
\end{align*}
We approximate ${k^0}^2$ as
\begin{align*}
{k^0}^2 &= (E_{p}-E_{p'})^2 \\
&\approx[\kappa_2 \epsilon \left(1+\frac{(\frac{\mathbf{q}}{\kappa_2})^2-\frac{2|\mathbf{b}|q_\parallel}{\kappa_2}}{2 \epsilon^2}\right)-\kappa_2\epsilon \left(1+\frac{(\frac{\mathbf{q'}}{\kappa_2})^2-\frac{2|\mathbf{b}|{q_\parallel}'}{\kappa_2}}{2 \epsilon^2}\right)]^2 \\
&=\left[ \frac{ |\mathbf{b}| (q_\parallel-{q_\parallel}')}{\epsilon}\right]^2+O(\alpha^3) \\
&\approx (\beta k_\parallel)^2.
\end{align*}
Using this, we find
\begin{align}
k^2 &\approx-(\mathbf{k}_\perp^2+\gamma^{-2}k_\parallel^2),
\end{align}
a result quoted earlier.

Inserting Eqs. (12) and (13) into (11) and doing the ${p'}^0$ integral gives
\begin{align}
 \frac{1}{\kappa_2}(\mathbf{q}_\perp^2+\gamma^{-2} q_\parallel^2-2\kappa_1 \kappa_2\Delta E \epsilon) f_1(p,b)&=  \frac{e^2}{(2\pi)^3}(\displaystyle{\not} b-\displaystyle{\not} p+m_1) \int  d^3p' \frac{1}{2E^{(2)}_{p'}} \frac{1}{\mathbf{k}_\perp^2+\gamma^{-2}k_\parallel^2}\notag \\
 & \;\;\;\;\times \gamma^\mu f_1(p',b) \gamma_\mu  (-\displaystyle{\not} p+m_2).
\end{align}
We next follow the analysis of \cite{Jarvinen} and approximate
\begin{align}
(\displaystyle{\not} b-\displaystyle{\not} p+m_1)\gamma^\mu f_1(p',b)\gamma_\mu (-\displaystyle{\not} p+m_2)&= 4E^{(2)}_p E^{(1)}_{b-p} \Lambda^+(\mathbf{b-p})\gamma^\mu f_1(p',b)\gamma_\mu \Lambda^-(\mathbf{p})\notag \\
&=4E^{(2)}_p E^{(1)}_{b-p} \frac{b^\mu}{\epsilon}f_1(p',b)\frac{b_\mu}{\epsilon} +O(\alpha) \notag \\
&=4E^{(2)}_p E^{(1)}_{b-p} \gamma^{-2}f_1(p',b).
\end{align}
where $\Lambda^{\pm}(\mathbf{p})=\frac{\pm \gamma^0 E_{\mathbf{p}} \mp \gamma \cdot \mathbf{p}+m}{2E_{\mathbf{p}}}$.  To lowest order in $\alpha$, we can write $E^{(1)}_{b-p}= \kappa_1 \epsilon$, $E^{(2)}_{p'}=E^{(2)}_{p}=\kappa_2 \epsilon$. Using Eq. (15), our final equation has the form
\begin{align}
 \frac{1}{\kappa_2}(\mathbf{q}_\perp^2+\gamma^{-2} q_\parallel^2-2\kappa_1 \kappa_2\Delta E \epsilon) f_1(p,b)&=  \frac{ 2 e^2 \epsilon \kappa_1}{\gamma^2 (2\pi)^3} \int  d^3p'  \frac{f_1(p',b)}{\mathbf{k}_\perp^2+\gamma^{-2}k_\parallel^2} \notag \\
 (\frac{1}{2\mu}(\mathbf{q}_\perp^2+\gamma^{-2} q_\parallel^2)- \Delta M) f_1(p,b)&=  \frac{e^2}{\gamma (2\pi)^3} \int  d^3p'  \frac{f_1(p',b)}{\mathbf{k}_\perp^2+\gamma^{-2}k_\parallel^2} 
\end{align}
where $\Delta M \equiv \gamma \Delta E$ and $\mu=\frac{m_1 m_2}{m_1+m_2}$.  The Lorentz contraction in the direction of motion is explicit in this equation.  Our result reduces to that of \cite{Jarvinen} in the equal mass case.

\section{Summary and future work}
We used the NQA to find an equation for a moving bound state consisting of two fermions of different masses.  After some approximations, we cast this equation into a form where the Lorentz contraction was evident.  This bound state equation matched that found through the Bethe-Salpeter procedure.  We did not elaborate much on the final answer, i.e. explain how to put the Dirac structure back in or show how the frame dependence can be removed through rescaling variables, because such things are well discussed in \cite{Jarvinen}.  The purpose of this paper was simply to promote an alternative method for arriving at the same answer.  Whether the N-Quantum was the simpler process in this case is debatable, but we feel it may be more useful in some other calculations.  Integrating over the mass shell delta-functions generated by the on shell lines connected to amplitudes should be easier than the alternative in many cases.  The N-Quantum also avoids the complications of the Bethe-Salpeter method discussed in the introduction.

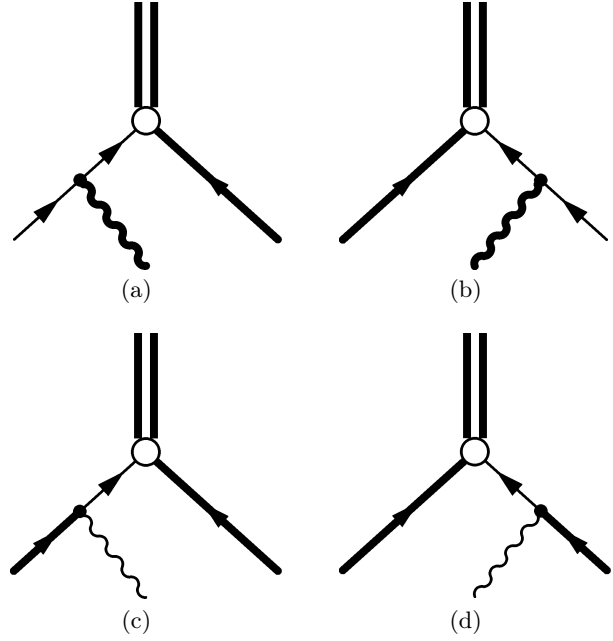
\begin{figure}[t]
\begin{center} $
\begin{array}{cc}
\subfloat[]{\label{fig:a}
\begin{fmffile}{graph1}
\begin{fmfgraph*}(100,100) \fmfpen{thin}
\fmfbottomn{i}{3} \fmftop{o1} 
\fmf{fermion,label=$1$}{i1,v1}
\fmf{fermion}{v1,v2}
\fmf{photon,width=25,tension=0}{i2,v1}
\fmf{fermion,width=25,label=$2$,tension=1/2}{i3,v2}
\fmf{double,width=25}{v2,o1}
\fmfdot{v1}
\fmfv{decor.shape=circle,decor.filled=empty,
decor.size=5thick}{v2}
\end{fmfgraph*}
\end{fmffile}}
&
\subfloat[]{\label{fig:b}
\begin{fmffile}{graph2}
\begin{fmfgraph*}(100,100) \fmfpen{thin}
\fmfbottomn{i}{3} \fmftop{o1} 
\fmf{fermion,label=$1$,width=25,tension=1/2}{i1,v2}
\fmf{fermion}{v1,v2}
\fmf{photon,width=25,tension=0}{i2,v1}
\fmf{fermion,label=$2$}{i3,v1}
\fmf{double,width=25}{v2,o1}
\fmfdot{v1}
\fmfv{decor.shape=circle,decor.filled=empty,
decor.size=5thick}{v2}
\end{fmfgraph*}
\end{fmffile}} \\
\subfloat[]{\label{fig:c}
\begin{fmffile}{graph3}
\begin{fmfgraph*}(100,100) \fmfpen{thin}
\fmfbottomn{i}{3} \fmftop{o1} 
\fmf{fermion,width=25,label=$1$}{i1,v1}
\fmf{fermion}{v1,v2}
\fmf{photon,tension=0}{i2,v1}
\fmf{fermion,width=25,label=$2$,tension=1/2}{i3,v2}
\fmf{double,width=25}{v2,o1}
\fmfdot{v1}
\fmfv{decor.shape=circle,decor.filled=empty,
decor.size=5thick}{v2}
\end{fmfgraph*}
\end{fmffile}}
&
\subfloat[]{\label{fig:d}
\begin{fmffile}{graph4}
\begin{fmfgraph*}(100,100) \fmfpen{thin}
\fmfbottomn{i}{3} \fmftop{o1} 
\fmf{fermion,width=25,label=$1$,tension=1/2}{i1,v2}
\fmf{fermion}{v1,v2}
\fmf{photon,tension=0}{i2,v1}
\fmf{fermion,width=25,label=$2$}{i3,v1}
\fmf{double,width=25}{v2,o1}
\fmfdot{v1}
\fmfv{decor.shape=circle,decor.filled=empty,
decor.size=5thick}{v2}
\end{fmfgraph*}
\end{fmffile}}
\end{array} $
\end{center}
\caption{Graphs for the matrix element $<B| \psi_{1}(p_1)\psi_{2}(p_2)A^\mu(k)|0>$.  The left fermion line is the electron and the right fermion line is the proton in each diagram.}
\end{figure}

In a subsequent paper, we will show how the N-Quantum can be used to calculate higher order Fock state contributions.  These terms can be expressed in terms of the lower order amplitudes with some approximation.  As an example, we can calculate the matrix element corresponding to the projection of the bound state onto the $ep\gamma$ Fock state, $M^\mu(p_1,p_2,k,\mathbf{b})\equiv \, <B| \psi_{1\alpha}(p_1)\psi_{2\beta}(p_2)A^\mu(k)|0>$.  After using the equations of motion to approximate some of the higher order amplitudes in terms of lower order ones, Haag expanding the interacting fields and contracting all \emph{in} fields, the result is
\begin{align}
&\frac{e}{(2 \pi)^3}\delta^4(p_1+p_2+k-b) \bigg\{ (-\displaystyle{\not} p_1+m_1) \gamma^\mu f_{1}(p_2,b)\bigg(\frac{\Theta_1(k,p_2)}{p_1^2-m_1^2}-\frac{\Theta_1(p_1,p_2)}{k^2}\bigg) \notag\\
& \;\;\;\;- C^T f_{2}^T(p_1,b)C\gamma^\mu(-\displaystyle{\not} p_2+m_2) \left(\frac{\Theta_1(k,p_1)}{p_2^2-m_2^2}-\frac{\Theta_1(p_1,p_2)}{k^2}\right)\bigg\}|_{b^0=E_{\mathbf{b}}}C
\end{align}
where $\Theta_1(k_i,k_j)=\frac{1}{6}\delta_{m_i}(k_i)\delta_{m_j}(k_j)(1+\theta(k_i^0)\theta(k_j^0)+\theta(-k_i^0)\theta(-k_j^0))$.  The $C$'s surround the transposed $f_2$ amplitude because the electron is arbitrarily given the "$\alpha$" index.  The results could have been left in terms of higher order amplitudes for a more exact, but more difficult to calculate solution.  We could also find this expression by interpreting the diagrams shown in figure (2) using the rules given in section (3).

It would be interesting to use the N-Quantum procedure to see whether or not classical Lorentz contraction takes place in higher order Fock state amplitudes.  This method can also be used to study other bound state models in which Lorentz covariance has been established.  The subject of Lorentz covariance and bound states in motion has been studied in a number of papers already \cite{Glockle} \cite{Glockle1} \cite{Artru}.  Some of the models within these works have interactions that lead to Lorentz invariant solutions \cite{Glockle1}, while others do not \cite{Artru}.  It would be interesting to analyze some of these models in the N-quantum framework, and determine which interactions result in Lorentz contracting solutions.  We hope that this paper has shown the utility of the N-quantum procedure in studying such models, and we plan to use it to gain a better understanding of bound state motion in future work.

\section*{Acknowledgements}
I would like to thank O. W. Greenberg for his guidance and helpful conversations.

\end{document}